# Study of the $^{20}$Ne(p,γ)$^{21}$Na reaction at LUNA

*Antonio* Caciolli[1,2,*] on behalf of the LUNA collaboration

[1]Dipartimento di Fisica ed Astronomia "G. Galilei", via Marzolo 8, 35131, Padova, Italy
[2]INFN Sezione di Padova, via Marzolo 8, 35131, Padova, Italy

**Abstract.** The NeNa-MgAl cycles are involved in the synthesis of Ne, Na, Mg, and Al isotopes. The $^{20}$Ne(p,γ)$^{21}$Na (Q = 2431.68 keV) reaction is the first and slowest reaction of the NeNa cycle and it controls the speed at which the entire cycle proceeds. At the state of the art, the uncertainty on the $^{20}$Ne(p,γ)$^{21}$Na reaction rate affects the production of the elements in the NeNa cycle. In particular, in the temperature range from 0.1 GK to 1 GK, the rate is dominated by the 366 keV resonance corresponding to the excited state of $E_X$ = 2797.5 keV and by the direct capture component. The present study focus on the study of the 366 keV resonance and the direct capture below 400 keV. At LUNA (Laboratory for Underground Nuclear Astrophysics) the $^{20}$Ne(p,γ)$^{21}$Na reaction has been measured using the intense proton beam delivered by the LUNA 400 kV accelerator and a windowless differential-pumping gas target. The products of the reaction are detected with two high-purity germanium detectors. The experimental details and preliminary results on the 366 keV resonance and on the direct capture component at very low energies will be shown, together with their possible impact on the $^{20}$Ne(p,γ)$^{21}$Na reaction rate.

## 1 Introduction

The aim of experimental Nuclear Astrophysics is to understand the nuclear processes involving in stars during their evolution and also to understand the mechanisms behind the Big Bang Nucleosynthesis [1]. In the last 30 year LUNA (Laboratory for Underground Nuclear Astrophysics [2]) has focused its work on studying charged particle induced reactions that occur in astrophysical scenarios. LUNA benefits of its favourable position in the underground national laboratory of Gran Sasso [3] where the background due to the cosmic rays is reduced by almost 5 orders of magnitude in the γ spectrum [4]. At energies below 3 MeV still the environmental background due to the radioisotopes is present. This can be reduced by an effective shielding composed by few cm of copper and about 25 cm of high purity lead [5, 6]. In this contribution, I will focus on the studies performed by LUNA on the Neon-Sodium
(NeNa) cycle in the last years with a particular attention on the recent results achieved on the $^{20}$Ne(p,γ)$^{21}$Na reaction. In the last section, I will also discuss the next challenges on what the LUNA activity is focused for the next years.

*e-mail: antonio.caciolli@unipd.it

## 2 The neon sodium cycle

The NeNa cycle converts hydrogen into helium utilizing neon and sodium isotopes as catalysts through the following reactions:

$$^{20}\text{Ne}(p,\gamma)^{21}\text{Na}(\beta^+\nu)^{21}\text{Ne}(p,\gamma)^{22}\text{Na}(\beta^+\nu)^{22}\text{Ne}(p,\gamma)^{23}\text{Na}(p,\alpha)^{20}\text{Ne}.$$

In Figure 1 the network of reactions involved in the NeNa cycle is depicted. In addition, the proton capture on the $^{23}$Na is shown opening to the production of Mg and Al isotopes.

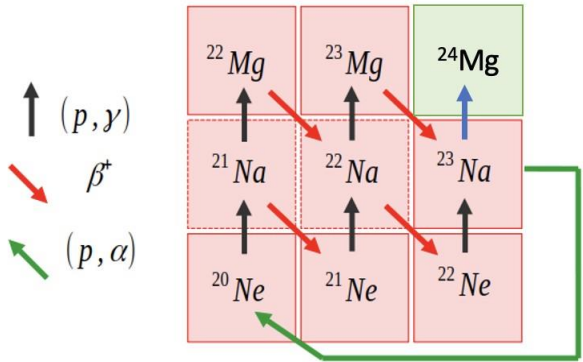

**Figure 1.** A schematic picture of the neon sodium cycle. The link to the magnesium aluminum one is also shown.

The products of NeNa nucleosynthesis process may become visible when transported to the stellar surface due to mixing with the stellar interior. Mixing occurs in asymptotic giant branch (AGB) stars with masses in the range of $M = 5–9\ M^\odot$, where the convective envelope extends into the hydrogen-burning layers, resulting in the transport of freshly synthesized material to the stellar surface. This phenomenon is known as Hot Bottom Burning (HBB) [7].

Consequently, the atmospheres of these massive AGB stars become enriched in nitrogen and sodium. Another astrophysical object affected by the NeNa cycle is ONe novae. Of particular interest is the 1275 keV γ-ray line associated with the $\beta^+$ decay of $^{22}$Na, which is essential for confirming a long-standing prediction of nova nucleosynthesis models [8].

Among the isotopes involved in the NeNa cycle, $^{20}$Ne is the most abundant, and the $^{20}$Ne$(p,\gamma)^{21}$Na reaction is the slowest in the cycle, significantly influencing the final abundances of Ne and Na isotopes. A sensitivity analysis regarding the impact of variations in the $^{20}$Ne$(p,\gamma)^{21}$Na rate on novae ejecta suggests a significant effect on isotopic abundances of elements with $A < 40$ [9].

Recently, the LUNA collaboration studied another important reaction on the NeNa cycle, the $^{22}$Ne$(p,\gamma)^{23}$Na reaction. Before LUNA results, this was the highest source of uncertainty due to the impact of several resonances where only upper limits due to indirect measurements were present [10, 11]. LUNA approached this reaction in two different campaigns: one characterised by the use of high resolution high purity germanium detectors [12] and the second where a 4π-BGO detector was implemented [13]. Those two experiments were able to measure for the first time three different resonances [14–16] and to provide direct capture data below 400 keV in the center of mass [17]. Additional important resonances were studied
on laboratory in surface [18] bringing to important impact on the nucleosynthesis in AGB models [19]. After this, we moved on the study of the $^{23}$Na($p,\gamma$)$^{24}$Mg reaction where three resonances of astrophysical interest were studied with an high efficiency BGO detector [20] with improved uncertainties with respect to the previous work in literature [21].

## 3 The $^{20}$Ne($p,\gamma$)$^{21}$Na Reactionreaction

At temperatures $T$ < 0.1GK, relevant for HBB, the $^{20}$Ne($p,\gamma$)$^{21}$Na reaction (Q-value = 2431.9keV) is primarily influenced by the high-energy tail of a sub-threshold state at $E_{cm}$ = −6.7keV ($\Gamma^\gamma$ = 0.31 ± 0.07eV), corresponding to the $E^x$ = 2425keV excited state in $^{21}$Na

[10]is governed by a narrow resonance at. At temperatures $T \overset{=}{\ } 0.1 – 1.0$GK, including those relevant to novae, the reaction rate$E_{cm}$ ≈ 366keV [22], corresponding to an excited state at $E_x$ = 2799keV in $_{21}$Na, as well as by direct capture contributions to the ground state and the first and second excited states in $^{21}$Na at $E_x$ = 332keV and 2425keV, respectively. The strength of the narrow resonance at $E_{cm}$ ≈ 366keV was first measured by Rolfs et al. [22] to ofbe$\omega\omega_{\gamma\gamma}$ == (0.11±0.02)0.0068)meV, while a recent study (Cooper, PhD thesis, [23]) reports a strengthmeV.

energy resonances, have also been documented in previous research ([24] and referenceDirect capture contributions at(0.0722 ± $E_{cm}$ ≥ 352keV, as well as contributions from highertherein). Notably, a non-resonant component was first explored in Ref. [25] using the activation method, employing the β$^+$ decay ($t_{1/2}$ = 22.4s) of $^{21}$Na into $^{21}$Ne at beam energies $E_{cm}$ = 600keV and $E_{cm}$ = 1050keV. Subsequently, a comprehensive study by Rolfs et al. investigated the direct component and several resonances at proton beam energies $E^{cm}$ = 352 – 2000keV, with the direct capture into the 2425keV sub-threshold state being dominant [22]. More recently, the $^{20}$Ne($p,\gamma$)$^{21}$Na reaction was indirectly examined using the $^{20}$Ne($^3$He,$d$)$^{21}$Na reaction [26]. The partial width of the sub-threshold state and the direct capture spectroscopic factors were calculated using the asymptotic normalization coefficient (ANC) formalism. The results align well with previous data for the direct capture to the 2425keV sub-threshold state, while a 65% discrepancy was noted for the direct capture into the ground state. New direct capture data were also recently reported by Lyons et al. at energiesIn the latter study, the direct capture and resonant components could not be clearly distin-guished, but the results for direct capture to the ground state were approximately 40% lower than those reported by Rolfs et al. Given the lack of low-energy data on direct capture and the high uncertainties associated with existing data [22, 23] on the $E_{cm} \approx 366$ keV strengths, improved measurements are essential to provide better constraints on the $_{20}$Ne$(p,\gamma)_{21}$Na reaction rate. $E_{cm}$ = 477 – 1905keV and by Karpesky (PhD thesis) [27] at energies $E_p < 400$ keV.

## 4 The experiment at LUNA

A setup similar to what described in [12] was installed on the windowless gas target beamline of the LUNA400 accelerator. In brief, two HPGe detectors were coupled to the scattering chamber in close geometry with a full shielding around made of few cm of high conductivity oxygen free copper and 25 cm of low counting lead. No internal collimators were present and natural neon gas (having about 90% of $^{20}$Ne inside) was inserted in the pumping system and kept at a pressure of 2 mbar. A typical beam current of 200 μA was used during the experiment. The beam current was determined with the calorimeter approach [28]. The beam heating effect [29] was taken into account as discussed in previous experimental campaigns [30]. Efficiency of detectors was calculated along the whole target length by using calibrated radioactive sources. A scheme of the setup together with the efficiency curves for one of the two detector are shown in Figure 2.

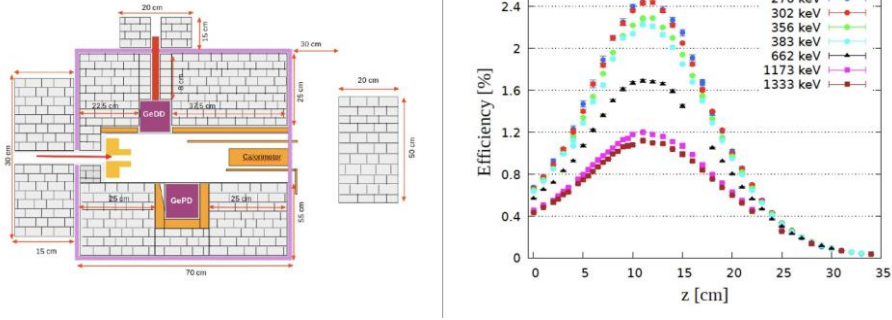

**Figure 2.** Left panel: a scheme of the setup. Right panel: the efficiency curves for all calibrated sources used in the experiment is shown for the detector called "GePD". An analogous plot is obtained for the other detector. The z=0 corresponds to the entrance aperture of the scattering chamber.

With this setup was possible to scan the resonance at the 366 keV on both detectors. The experimental yield obtained on both detectors is depicted in Figure 3. Thanks to the high precision of the data we were able to recover the resonance strength on each point of both scans finding values with a very perfect agreement. The final ωγ value has been found in good agreement with the Rolfs one but with improved precision of a factor of 10 in the statistical part and on a factor of 5 in the systematic component. The scans have been fitted with a skewed Gaussian function. Since the maximum of each correspond to the population of the resonance in front of the detector, by knowing the energy loss in the gas and the gas target density in the scattering chamber, the last collimator and the beampipe before entering in the chamber, it is possible to also determine the value of the resonance energy. This is an important parameter that enters in the reaction rate calculation. The value found is $E_R^{cm}$ = 368.0 ± 0.5 off by 2 keV with respect to the value in [22] but still in agreement with that work due to their large error bars. In addition to that experimental data points for the non resonant component of the cross section have been obtained in the energy region below 400 keV. In this energy region no published data points are present up to now. Those values constraint the extrapolations at low energies having an effect of reducing the reaction rate by 20% with respect the work of Lyons et al. [24].

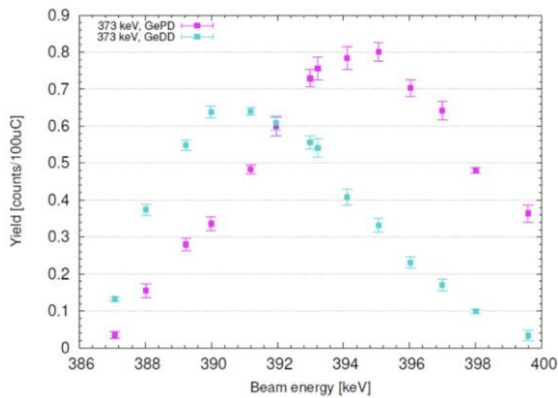

**Figure 3.** Resonance scans of the 366 keV resonance on both detectors. The maximum peak is shifted since the position of each detector is different with respect to the entrance of the scattering chamber.

The work of LUNA on the NeNa cycle is continuing in the end of 2023 studying the $^{21}$Ne(p,γ)$^{22}$Na reaction with the same setup discussed here. In the last months of 2023 a new setup will be mounted on the LUNA400 accelerator involving silicon detectors to study the $^{23}$Na(p,α)$^{20}$Ne reaction. The data taking is expected to start in Q1 2024. It is worth to mention that on the solid target beamline new exciting results have been recently obtained on several reaction of the CNO cycles such as the proton captures on carbon isotopes [31, 32] or the proton capture on $^{17}$O where we are going to extend previous works [33–35] studying the low resonance at 65 keV with a high impact in stellar nucleosynthesis [36].

## 5 A new MV accelerator recently installed at LNGS

A new MV machine, capable of accelerate high precision and high intensity proton, helium and carbon beams, has been installed at the LNGS. This machine could open to the possibility of studying the helium and carbon burnings. LUNA has proposed some scientific cases to be studied with this machine in the next years. In particular, the first reaction planned to the tackled is the $^{14}$N(p,γ)$^{15}$O where there are many studied in literature but still a large discrepancy in the extrapolations at solar energies (see [37] and reference therein). This reaction is important to the calculation of Solar neutrinos and can be used to probe the solar metallicity by combining nuclear inputs and the recent experimental CN neutrino observations from Borexino and also the one that will be provided by future experiments.

During 2023 and 2024 another reaction will be studied with this machine, the $^{22}$Ne(α,n)$^{25}$Mg reaction. This is one of the neutron emitters for the s-process. Finally, the $^{12}$C+$^{12}$C will be studied by the start of 2024. A high efficiency and low background γ-ray setup will be installed at one of the beamline of the new MV machine. This setup will grant an high sensitivity allowing to measure the γ rays emitted by the proton and alpha channels of this reaction down to energies below 2 MeV. With those results LUNA could constrain the discrepancy in the various experiments in literature granting a reliable value for the M$_{up}$ parameter. This is important since this parameter describe which is the threshold mass value to ignite the carbon burning and affects the fate of stellar objects.

## 6 Conclusion

In this contribution new results on the $^{20}$Ne(p,γ)$^{21}$Na reaction studied at the LUNA400 accelerator have been reported along with the description of the setup used. With this accelerator further important studies on the NeNa cycle will be done in next future, but addition scientific cases are planned for the next years. At the same time the presence of a new MV machine installed at the LNGS will provide the unique opportunity to study helium and carbon burning in the next years.

## Acknowledgments

D. Ciccotti and the technical staff of the LNGS are gratefully acknowledged for their support. We acknowledge funding from: INFN, the Italian Ministry of Education, University and Research (MIUR) through the "Dipartimenti di eccellenza" project "Physics of the Universe", and ChETEC-INFRA, no. 101008324), the European Collaboration for Science and Technology (COST Action ChETEC, CA16117). For the purpose of open access, authors have applied a Creative Commons Attribution (CC BY) licence to any Author Accepted Manuscript version arising from this submission.